\newif\ifpdf
\ifx\pdfoutput\undefined
\pdffalse % we are not running PDFLaTeX
\else
\pdfoutput=1 % we are running PDFLaTeX
\pdftrue \fi

\documentclass[reqno,twoside,11pt]{amsart}
\usepackage{cite}
\usepackage{amsmath}
\usepackage{amsfonts}
\usepackage{amssymb}
\usepackage{wrapfig}
\usepackage{graphicx}
\usepackage{graphics}
\usepackage{verbatim}
\usepackage{xr}
\usepackage{exscale}

\IfFileExists{epsf.def}{\input epsf.def}{\usepackage{epsf}}

\IfFileExists{myowntimes.sty}{\usepackage{myowntimes}}
{\usepackage{times}}
\IfFileExists{mathrsfs.sty}{\usepackage{mathrsfs}}{\newcommand{\mathscr}{\mathcal}}
\DeclareFontFamily{OT1}{eufm}{} \DeclareFontShape{OT1}{eufm}{m}{n}
{<5> <6> <7> <8> <9> <10> <11> <12> <14.4> eufm10}{}
\DeclareMathAlphabet{\mathfrak}{OT1}{eufm}{m}{n}

\DeclareFontFamily{OT1}{cmss}{} \DeclareFontShape{OT1}{cmss}{m}{n}
{<5> <6> <7> <8> <9> <10> <11> <12> <13> <14.4> cmss10}{}
\DeclareMathAlphabet{\cmss}{OT1}{cmss}{m}{n}
\newtheoremstyle{thm}{1.5ex}{1.5ex}{\itshape\rmfamily}{}
{\bfseries\rmfamily}{}{2ex}{}

\newtheoremstyle{rem}{1.3ex}{1.3ex}{\rmfamily}{}
{\itshape}
{} {1.5ex}{}

\newenvironment{proofsect}[1]
{\vskip0.1cm\noindent{\rmfamily\itshape#1.}}{\qed\vspace{0.15cm}}

\theoremstyle{thm}
\newtheorem{theorem}{Theorem}[section]
\newtheorem{lemma}[theorem]{Lemma}

\newtheorem*{Main Theorem}{Main Theorem.}

\theoremstyle{rem}
\newtheorem{remark}{{\itshape Remark}}[]

\numberwithin{equation}{section}

\renewcommand{\section}{\secdef\sct\sect}
\newcommand{\sct}[2][default]{\refstepcounter{section}
\addcontentsline{toc}{section}
{{\tocsection {}{\thesection}{\!\!\!\!#1\dotfill}}{}}
\vspace{0.7cm}
\centerline{ %\large
\scshape\arabic{section}.\ #1} \nopagebreak \vspace{0.2cm}}
\newcommand{\sect}[1]{
\vspace{0.4cm} \centerline{\large\scshape\rmfamily #1}
\vspace{0.2cm}}

\renewcommand{\subsection}{\secdef\subsct\sbsect}
\newcommand{\subsct}[2][default]{\refstepcounter{subsection}
\addcontentsline{toc}{subsection}
{{\tocsection{\!\!}{\hspace{1.2em}\thesubsection}{\!\!\!\!#1\dotfill}}{}}
\nopagebreak \vspace{0.45\baselineskip} {\flushleft\bf
\arabic{section}.\arabic{subsection}~\bf #1.~}
\\*[3mm]\noindent
\nopagebreak}
\newcommand{\sbsect}[1]{\vspace{0.1cm}\noindent
\textbf{#1.~}\vspace{0.1cm}}

\renewcommand{\subsubsection}{%
\secdef \subsubsect\sbsbsect}
\newcommand{\subsubsect}[2][default]{%
\refstepcounter{subsubsection} 
\addcontentsline{toc}{subsubsection}{{\tocsection{\!\!}
{\hspace{3.05em}\thesubsubsection}{\!\!\!\!#1\dotfill}}{}}
\nopagebreak
\vspace{0.15\baselineskip} \nopagebreak {\flushleft\rmfamily
\itshape\arabic{section}.\arabic{subsection}.\arabic{subsubsection}
\ \rmfamily #1\/.}\ }
\newcommand{\sbsbsect}[1]{\vspace{0.1cm}\noindent
\rmfamily \itshape
\arabic{section}.\arabic{subsection}.\arabic{subsubsection} \
\sffamily #1\/.\ }

\newcommand{\dist}{\operatorname{dist}}

\newcommand{\diam}{\operatorname{diam}}

\newcommand{\textd}{\text{\rm d}}

\renewcommand{\AA}{\mathcal A}
\newcommand{\BB}{\mathcal B}

\newcommand{\A}{\mathbb A}

\newcommand{\BbbH}{\mathbb H}

\newcommand{\BbbL}{\mathbb L}

\newcommand{\BbbP}{\mathbb P}

\newcommand{\R}{\mathbb R}
\newcommand{\BbbS}{\mathbb S}

\newcommand{\Z}{\mathbb Z}
\newcommand{\bn}{{\boldsymbol n}}

\newcommand{\Deltac}{{\Delta_{\text{\rm c}}}}

\newcommand{\betac}{{\beta_{\text{\rm c}}}}
\newcommand{\rhoG}{\rho_{\text{\rm g}}}
\newcommand{\rhoL}{\rho_{\lower1.2pt\hbox{\fontsize{8}{8}\selectfont$\ell$}}}
\newcommand{\barrhoG}{{\bar\rho_{\text{\rm g}}}}
\newcommand{\ZC}{Z_{\text{\rm C}}}
\newcommand{\ZG}{Z_{\text{\rm G}}}
\newcommand{\mut}{\mu_{\text{\rm t}}}

\newcommand{\ssP}{\cmss P}
\newcommand{\ssC}{\cmss C}
\newcommand{\mstar}{m^\star}

\newcommand{\ext}{{\text{\rm ext}}}

\renewcommand{\eqref}[1]{(\ref{#1})}
\newcommand{\twoeqref}[2]{(\ref{#1}--\ref{#2})}

\newcommand{\tV}{\mathchoice%
	{\text{\fontsize{8}{5}\selectfont\hbox{$\pmb\triangle$}\hskip-1pt}V}
	{\text{\fontsize{8}{5}\selectfont\hbox{$\pmb\triangle$}\hskip-1pt}V}
	{\text{\fontsize{5}{4}\selectfont\hbox{$\pmb\triangle$}\hskip-1pt}V}
	{\text{\fontsize{5}{4}\selectfont\hbox{$\pmb\triangle$}\hskip-1pt}V}}
\newcommand{\deltaV}{{\delta\mkern-1mu V}}
\newcommand{\deltaN}{{\delta\mkern-1mu N}}
\newcommand{\pV}{p_{\scriptscriptstyle V}}
\newcommand{\pL}{p_{\scriptscriptstyle L}}
\newcommand{\pinfty}{p_{\scriptscriptstyle\infty}}

\begin{document}

\title[A microscopic theory of Gibbs-Thomson formula]
{\Large A proof of the Gibbs-Thomson formula\\ in the droplet formation regime}
\author[M.~Biskup, L.~Chayes and R.~Koteck\'y]
{Marek~Biskup,${}^1\,$ Lincoln~Chayes${}^1$ and$\,$
Roman~Koteck\'y${}^2$}
%\begin{abstract}
%\end{abstract}

\thanks{\hglue-4.5mm\fontsize{9.6}{9.6}\selectfont\copyright\,\,\,2003 by the authors. Reproduction, by any means, of the entire article for non-commercial purposes is permitted without~charge.\vspace{2mm}}
\maketitle

\vspace{-4mm} \centerline{${}^1$\textit{Department of Mathematics,
UCLA, Los Angeles, California, USA}} %\vspace{-4mm}
\centerline{${}^2$\textit{Center for Theoretical Study, Charles
University, Prague, Czech Republic}}

\vspace{3mm}
\begin{quote}
\footnotesize {\bf Abstract:} 
We study equilibrium droplets in two-phase systems at parameter values corresponding to phase coexistence. 
Specifically, we give a self-contained microscopic derivation of the Gibbs-Thomson formula for the deviation 
of the pressure and the density away from their equilibrium values which, according to the interpretation 
of the classical thermodynamics,  appears due to the presence of a curved interface. 
The general---albeit heuristic---reasoning is corroborated by a rigorous proof in the case 
of the two-dimensional~Ising lattice gas.
\end{quote}
\vspace{2mm}

\section{Introduction}
%\smallskip\noindent
\subsection{The problem}
\label{sec1.1}\noindent
The description of equilibrium droplets for systems with coexisting phases is one of the 
outstanding achievements of classical thermodynamics. 
Standard treatments of the subject 
highlight various formul\ae{}
relating the linear size of the droplet to a specific pressure difference.
One of these, called the \emph{Gibbs-Thomson} formula, concerns the difference between the actual pressure 
outside the droplet and the ambient pressure of the system without any droplets.
(Or, in the terminology used in classical textbooks, ``above a curved interface'' 
and ``above a planar interface,'' respectively.) 
The standard reasoning behind these formul\ae{} is based primarily 
on macroscopic concepts of pressure, surface tension, etc. 
But, notwithstanding their elegance and simplicity,
these derivations do not offer much insight into the microscopic aspects of droplet equilibrium. 
The goal of the present paper is to give a self-contained derivation 
of the Gibbs-Thomson formula starting from the first principles of \emph{equilibrium} statistical mechanics. 

While straightforward on the level of  macroscopic thermodynamics,
an attempt for a microscopic theory of droplet equilibrium immediately reveals several 
technical problems. 
First of all, there is no obvious way---in equilibrium---to discuss finite-sized droplets
that are immersed in an \emph{a priori} infinite system. Indeed, the correct setting is the asymptotic behavior 
of finite systems that are scaling to infinity and that contain droplets whose size \emph{also} scales to infinity 
(albeit, perhaps, at a different rate).
Second, a statistical ensemble has to be produced  whose typical configurations will feature an equilibrium droplet 
of a given linear size. A natural choice is the canonical ensemble with a tiny fraction of extra  
particles tuned so that a droplet of a given size is induced in the system.
A difficulty here concerns the existence of a minimal droplet size as will be detailed below.
Finally, for the specific problem at hand, the notions of pressure ``above a curved interface'' 
and ``above a planar interface'' have to be reformulated in terms of microscopic quantities 
which allow for a comparison of the difference between these pressures and the droplet~size.

Some of these issues have previously been addressed by the present authors. Specifically, in~\cite{BCK,bigBCK}, 
we studied the droplet formation/dissolution phenomena in the context of the canonical ensemble 
at parameters corresponding to phase coexistence and the particle density slightly exceeding
the ambient limiting rarefied density. 
It was found that, if $V$ is the volume of the system and~$\deltaN$ is the particle excess, 
droplets form when the ratio~$(\deltaN)^{(d+1)/d}/V$ is of the order of unity. 
In particular, there exists a dimensionless parameter~$\Delta$, 
proportional to the thermodynamic limit of this ratio, 
and a non-trivial critical value~$\Deltac$, such that, for $\Delta<\Deltac$, 
all of the excess will be absorbed into the (Gaussian) fluctuations of the ambient gas, 
while if $\Delta>\Deltac$, a mesoscopic droplet will form. 
Moreover, the droplet will only subsume a fraction~$\lambda_\Delta<1$ of the excess particles.
This fraction gets smaller as $\Delta$ decreases to $\Deltac$, yet the minimum fraction~$\lambda_{\Deltac}$ 
does \emph{not} vanish. 
It is emphasized that these minimum sized droplets are a mesoscopic phenomenon:
The linear size of the droplet will be proportional to~$V^{1/(d+1)}\ll V^{1/d}$ and the droplet thus occupies 
a vanishing fraction of the system. Note that the total volume cannot be 
taken arbitrary large 
if there is to be a fixed-size droplet~at~all.

The droplet formation/dissolution phenomena have been the subject of intensive study in last few years. 
The fact that~$d/(d+1)$ is the correct \emph{exponent} for the scale on which droplets first appear 
was shown rigorously in~\cite{DS} (see also~\cite{Bob+Tim}); a heuristic derivation may go back 
at least to~\cite{Binder-Kalos}. The existence of a \emph{sharp} minimal droplet size on the scale $V^{1/(d+1)}$ 
was described in~\cite{Sethna}, more recently in~\cite{BCK,Neuhaus-Hager} and yet again in~\cite{Binder}. 
In the context of the 2D~Ising system, a rigorous justification of the theory outlined in~\cite{BCK} 
was provided in~\cite{bigBCK}. We note that the existence of a minimal droplet size seems to be ultimately related 
to the pressure difference ``due'' to the presence of a droplet as expressed by the Gibbs-Thomson formula. 
Indeed, from another perspective (which is more or less that of~\cite{Sethna,Neuhaus-Hager}), 
the formation/dissolution phenomena can be understood on the basis of arguments in which the Gibbs-Thomson formula 
serves as a foundation.
Finally, we remark that although the generation of droplets is an inherently dynamical phenomenon
(beyond the reach of current methods) it is possible that, on limited temporal and spatial scales, the equilibrium asymptotics is of direct relevance.

The remainder  of this paper is organized as follows. In the next subsection (Section~\ref{sec1.2}) 
we will present an autonomous derivation of the Gibbs-Thomson formula based on first principles 
of statistical mechanics. 
Aside from our own (modest) appreciation of this approach, Section~\ref{sec1.2} is worthwhile in the present context 
because the rigorous analysis develops precisely along these lines. 
In Section~\ref{sec2},  we will restrict our attention to the 2D~Ising lattice gas, define explicitly 
the relevant quantities and present our rigorous claims in the form of mathematical theorems. 
The proofs will come in Section~\ref{sec3}. 

\subsection{Heuristic derivation}
\label{sec1.2}\noindent
Let us consider a two-phase system at parameter values corresponding to phase coexistence.
We will assume that the two phases are distinguished by their densities and, although the forthcoming
derivation is completely general, we will refer to the dense phase as \emph{liquid} and to the rarefied phase 
as \emph{gas}. 
Confining the system to a ($d\ge 2$)-dimensional volume~$V$, we will consider a canonical ensemble at
inverse temperature~$\beta$ and the number of particles fixed to the value 
\begin{equation}
\label{N}
N=\rhoG V + (\rhoL - \rhoG)\deltaV.
\end{equation}
Here,~$\rhoL$ and $\rhoG$ are the bulk densities of the liquid and gas, respectively, 
and the particle excess is~$\delta N= (\rhoL - \rhoG)\deltaV$ with $\deltaV\ll V$.
Let~$w_1$ denote the dimensionless  interfacial free energy (expressed in multiples of~$\beta^{-1}$), 
which represents the 
cost of an optimally-shaped droplet of unit volume, and let~$\varkappa$ denote 
the response function, $\varkappa= \frac1V \langle (N-\langle N\rangle)^2\rangle$, which is essentially 
the isothermal compressibility.
Then, as has been argued in~\cite{BCK}, if the parameter
\begin{equation}
\label{Delta}
\Delta=\frac{(\rhoL-\rhoG)^2}{2\varkappa w_1}\frac{(\deltaV)^{\frac{d+1}d}}{V},
\end{equation}
is less than a critical value~$\Deltac=\frac1d (\frac{d+1}2 )^{\frac{d+1}d}$,
all of the particle excess will be absorbed by the background fluctuations, while, for $\Delta>\Deltac$,
a fraction of the excess particles will condense into a droplet. Moreover, the volume of this droplet will be 
(in the leading order)~$\lambda_\Delta\deltaV$, where $\lambda_\Delta\in[0,1]$ is the maximal solution to the equation 
\begin{equation}
\label{eqlambda}
\frac{d-1}d \lambda^{-1/d} = 2\Delta (1-\lambda).
\end{equation}
Note that $\lambda_{\Deltac}=2/(d+1)$ as advertised; that is to say, the droplet does not appear gradually. 
Furthermore,  as is of interest in  certain anisotropic situations
where the droplet plays a role of an equilibrium crystal, 
the droplet has a particular shape, 
known as the \emph{Wulff shape}, which optimizes the overall interfacial free energy for a given volume.

\subsubsection{Gibbs-Thomson~I: The density}
On the basis of the aforementioned claims, we can already state a version of the Gibbs-Thomson formula 
for the difference of densities ``due to the presence of a curved interface.''
Indeed, since the droplet only accounts for a fraction, $\lambda_\Delta$, of the excess particles, 
the remainder~$(1-\lambda_\Delta) (\rhoL-\rhoG)\deltaV$, of these particles reside in the bulk.
Supposing that the droplet subsumes only a negligible fraction of the entire volume, i.e.,~$\deltaV\ll V$, 
the gas surrounding  the droplet will thus have the density
\begin{equation}
\label{barrhoA}
\barrhoG=\rhoG+(1-\lambda_\Delta) (\rhoL-\rhoG)\frac{\deltaV}V\bigl(1+o(1)\bigr).
\end{equation}
Here~$o(1)$ is a quantity tending to zero as~$V$ tends to infinity while keeping~$\Delta$ finite
(and~$\Delta>\Deltac$).
Invoking~\eqref{Delta} and~\eqref{eqlambda}, this is easily converted into
\begin{equation}
\label{barrhoB}
\barrhoG=\rhoG+\frac{d-1}d \frac{\varkappa w_1}{\rhoL-\rhoG}\frac{1}{(\lambda_\Delta \deltaV)^{1/d}}\bigl(1+o(1)\bigr).
\end{equation}
Thus, the density of the gas surrounding the droplet will exceed the density of the ambient gas by a factor
inversely-proportional to the linear size of the droplet. 
This is (qualitatively) what is stated by the Gibbs-Thomson formula.

In order to 
make correspondence with physics literature,
let us assume that the droplet is spherical---which 
is the case for an isotropic surface tension. 
Then we have
\begin{equation}
%\label{}
w_1=\beta\sigma\, S_d \Bigl(\frac{S_d}d\Bigr)^{-\textstyle\frac{d-1}d}
\quad\text{and}\quad
\lambda_\Delta \deltaV=\frac{S_d}d r^d
\end{equation}
where~$\sigma$ is the surface tension,~$S_d$ is the surface area of a unit sphere in~$\R^d$ and~$r$ is
the radius of the droplet.
Substituting these relations into~\eqref{barrhoB}, we will get
\begin{equation}
\label{barrhoC}
\barrhoG=\rhoG+(d-1)\frac{\beta\sigma\varkappa}{\rhoL-\rhoG}\frac1r\bigl(1+o(1)\bigr).
\end{equation}
Of course, all three formulas \eqref{barrhoA}, \eqref{barrhoB} and \eqref{barrhoC} represent
the leading order asymptotic in~$1/r$. 
Higher-order corrections go beyond the validity of the presented argument.

\begin{remark}
\label{rem1}
We note that equation~\eqref{barrhoC} differs from the usual corresponding version of the Gibbs-Thomson formula 
in which the~$\varkappa$ appearing above is replaced by~$\rhoG$. 
This is due to the approximation
$\varkappa\approx\rhoG$
which is justified only in the ideal-gas limit 
of the rarefied phase.
\end{remark}

\subsubsection{Pressures above curved/planar interfaces}
Next we turn our attention to the Gibbs-Thomson formula for the pressure.
Here we immediately run into a complication; while the density is a well-defined object in finite volume, 
the pressure, by its nature, is a macroscopic commodity.
Thus, strictly speaking, the \emph{pressure} should be discussed in the context of thermodynamic limits.

In the present context we need to define the ``pressure of the gas surrounding a droplet.'' In order to do so, 
we will consider two canonical ensembles with the \emph{same} number of particles given by~\eqref{N}, 
in volumes~$V$ and $V+\tV$, where~$\tV\ll V$. From the perspective of equilibrium thermodynamics, 
these two situations describe the initial and terminal states of the gas undergoing isothermal expansion.
Standard statistical-mechanical formulas tell us that the change of the relevant thermodynamic potential
(the Helmholtz free energy) during this expansion is given as the pressure times the difference of the volumes~$\tV$.
Using~$\ZC(N,V)$ to denote the \emph{canonical} partition function of~$N$ particles 
in volume~$V$, we thus define the relevant pressure~$\pV$ by
\begin{equation}
\label{pV}
\pV=\frac1\beta\frac1{\tV}\log\frac{\ZC(\rhoG V+(\rhoL-\rhoG)\deltaV,V+\tV)}{\ZC(\rhoG V+(\rhoL-\rhoG)\deltaV,V)}.
\end{equation}
For finite~$V$, $\tV$, etc., the quantity~$\pV$ still depends on~$\tV$. 
As it turns out, this
dependence (which we will refrain from making notationally explicit) will annul in any limit $V,\tV\to\infty$ with 
$\tV/\partial V\to0$, where~$\partial V$ denotes the boundary of~$V$.
However, we must consider a limiting procedure for which~$\tV$ also does not ``disturb'' the droplet. 
This is a slightly delicate subject matter to which we will return shortly.

Our next goal is to give a mathematical interpretation of the pressure ``above a planar interface.''
As it turns out (and as is the standard in all derivations), here the correct choice is to take simply the pressure 
of the ambient gas phase. (See Remark~\ref{rem2a} for further discussions.)
Using~$\ZG(\mu,V)$  to denote the \emph{grand canonical} partition function,  with~$\mu$ denoting 
the chemical potential, this quantity is defined by the (thermodynamic) limit
\begin{equation}
\label{therm-lim}
\pinfty=\frac1\beta\lim_{V\to\infty}\frac1V\log
\ZG(\mut,V),
\end{equation}
Here we have prepositioned the chemical potential to the transitional value, i.e.~$\mu=\mut$.
By well-known arguments, this limit is independent of how~$V$ tends to infinity provided~$\partial V/V$
tends to zero as~$V\to\infty$.

Since we are ultimately 
looking for an expression for the difference~$\pV-\pinfty$, instead of~\eqref{therm-lim} we would
rather have an expression that takes a form similar to~\eqref{pV}. 
We might try to use the fact that~$\log\ZG(\mut,V)=\beta \pinfty V+O(\partial V)$, 
but then the boundary term  will be much larger than the actual Gibbs-Thomson correction. 
We thus have to develop a more precise representation of the grand canonical partition function.
For simplicity, we will restrict ourselves to the cases when~$V$ is a rectangular box, in which case we expect to have
\begin{equation}
\label{ZGrep}
\log \ZG(\mut,V)=\beta \pinfty V+\tau_{\text{wall}} \partial V + O\bigl(V^{\frac{d-2}d}\bigr).
\end{equation}
Here~$\tau_{\text{wall}}$ denotes a \emph{wall surface tension} which depends on the boundary conditions.
The error term represents the contribution from lower-dimensional facets of~$V$, e.g., 
edges and 
corners of~$V$ in~$d=3$.
Such a representation can be justified using low-temperature expansions, see~\cite{Borgs-Kotecky}, 
and/or by invoking rapid decay of correlations. 
Of course, this will be discussed in excruciating detail  in Section~\ref{sec3} of the present paper.

Using the representation~\eqref{ZGrep}, we can now write
\begin{equation}
\label{pinf}
\beta \pinfty=\frac1{\tV}\log\frac{\ZG(\mut,V+\tV)}{\ZG(\mut,V)}+
O\Bigl(\frac{\partial (V+\tV)-\partial V+V^{\frac{d-2}d}}{\tV}\Bigr),
\end{equation}
which supposes that both~$V$ and~$V+\tV$ are rectangular volumes.

Our goal is to limit~$\tV$ to the values for which the error term is negligible compared
with the anticipated Gibbs-Thomson correction.
First, supposing that~$\tV\ll V$, we find that the difference~$\partial (V+\tV)-\partial V$ is 
of the order~$\tV/V^{1/d}$.
Second, assuming that~$\Delta$ from~\eqref{Delta} is finite and exceeding~$\Deltac$ (which is necessary to have any
droplet at all),  we have~$\deltaV\sim V^{d/(d+1)}$.
These two observations show that the contribution of~$\partial (V+\tV)-\partial V$ to the error term in~\eqref{pinf} is
indeed negligible compared with~$(\deltaV)^{-1/d}$. 
A similar calculation shows that the 
the second part of the error term,~$V^{(d-2)/d}/\tV$, on the right-hand side of~\eqref{pinf} 
is negligible compared with~$(\deltaV)^{-1/d}$ provided that
\begin{equation}
\label{tVcond}
\tV\gg V^{\frac{d-2}d+\frac1{d+1}}.
\end{equation}
It is easy to check---see formula~\eqref{tVagain}---that~\eqref{tVcond} can be satisfied while 
maintaining~$\tV\ll\deltaV$. This observation will be essential in the forthcoming developments.

The formulas \twoeqref{pV}{pinf} can be conveniently subtracted in terms of the probability~$\BbbP_V(N)$ that,
in the grand canonical ensemble, there are \emph{exactly}~$N$ particles in volume~$V$.
Explicitly, denoting
\begin{equation}
\label{PV}
\BbbP_V(N)=\frac{e^{\beta\mut N}\ZC(N,V)}{\ZG(\mut,V)},
\end{equation}
we get
\begin{equation}
\label{pV-pinf}
\beta(\pV- \pinfty)=\frac1{\tV}\log\frac{\BbbP_{V+\tV}(\rhoG V+(\rhoL-\rhoG)\deltaV)}
{\BbbP_V(\rhoG V+(\rhoL-\rhoG)\deltaV)}+o\bigl((\deltaV)^{-1/d}\bigr).
\end{equation}
Here we have applied~\eqref{tVcond} to simplify the error term.

\subsubsection{Gibbs-Thomson~II: The pressure}
Now we are in a position to derive the desired Gibbs-Thomson formula for the pressure.
A principal tool for estimating the ratio of the probabilities in~\eqref{pV-pinf} will be
another result of~\cite{BCK} which tells us that, in the limit~$V\to\infty$, 
\begin{equation}
\label{PV1}
-\log \BbbP_V\bigl(\rhoG V+ (\rhoL-\rhoG)\deltaV\bigr)=w_1 (\deltaV)^{\frac{d-1}d}
\bigl(\Phi^\star_\Delta+o(1)\bigr),
\end{equation}
where~$\Phi^\star_\Delta$ is the absolute minimum of the function
\begin{equation}
\label{Phi}
\Phi_\Delta(\lambda)=\lambda^{\frac{d-1}d}+\Delta(1-\lambda)^2
\end{equation}
on~$[0,1]$. Since~$\rhoG V+ (\rhoL-\rhoG)\deltaV=\rhoG(V+\tV)+ (\rhoL-\rhoG)(\alpha\,\deltaV)$, where
\begin{equation}
%\label{}
\alpha = 1-\frac{\rhoG}{\rhoL-\rhoG}\frac{\tV}{\deltaV}
\end{equation}
we also have, again in the limit~$V\to\infty$, 
\begin{equation}
\label{PV2}
-\log \BbbP_{V+\tV}\bigl(\rhoG V+ (\rhoL-\rhoG)\deltaV\bigr)=w_1 (\alpha\,\deltaV)^{\frac{d-1}d}
\bigl(\Phi^\star_{\Delta(\alpha)}+o(1)\bigr),
\end{equation}
where we have introduced the shorthand
$\Delta(\alpha)= \alpha^{\frac{d+1}d} \Delta$. 

Supposing that~$\tV\ll \deltaV$, we can write
\begin{equation}
%\label{}
\Phi^\star_{\Delta(\alpha)}=\Phi^\star_{\Delta}-\frac{\rhoG}{\rhoL-\rhoG}\frac{\tV}{\deltaV}
(1-\lambda_\Delta)^2+o(\tV/\deltaV)
\end{equation}
and thus, to the leading order in~$\tV/\deltaV$,
\begin{equation}
\label{1.15}
\beta(\pV- \pinfty)=w_1\frac{\rhoG}{\rhoL-\rhoG}\frac1{(\deltaV)^{1/d}}
\biggl[\frac{d-1}d \Phi^\star_\Delta + \frac{d+1}d \Delta (1-\lambda_\Delta)^2+o(1)\biggr].
\end{equation}
After some manipulations involving~\eqref{Phi} and~\eqref{eqlambda}, 
the square bracket on the right-hand side turns out to equal $\frac{d-1}d \lambda_\Delta^{-1/d}+o(1)$. Thus we finally derive
\begin{equation}
\label{1.17}
\beta(\pV- \pinfty)=\frac{d-1}d\frac{w_1\rhoG}{\rhoL-\rhoG}\frac1{(\lambda_\Delta\deltaV)^{1/d}}
\bigl(1+o(1)\bigr).
\end{equation}
In the case of an isotropic surface tension, formula~\eqref{1.17} again reduces to
\begin{equation}
\label{1.17b}
\pV- \pinfty=(d-1)\frac{\sigma\rhoG}{\rhoL-\rhoG}\frac1r
\bigl(1+o(1)\bigr).
\end{equation}
This is the (leading order) Gibbs-Thomson correction; the one which is usually derived~\cite{Landau,Tabor} 
by invoking \emph{thermodynamic} considerations.
We note that here the gas-density~$\rhoG$ in the numerator is fully justified, cf~Remark~\ref{rem1}.

\begin{remark}
\label{rem2}
We note that higher orders in~$1/r$---as predicted by the ``exponential'' Gibbs-Thomson formula 
in classical thermodynamics---go beyond the validity of the formulas~\eqref{PV1} and~\eqref{PV2}. 
In fact, as a closer look at the~$V$-dependence of~$\deltaV$ and~$\partial V$ suggests, 
these corrections may depend on the choice of the volumes~$V$ and~$V+\tV$ and on the boundary condition.
We further remark that both formulas~\eqref{barrhoB} and~\eqref{1.17} have been derived 
for the situation when a droplet of the dense phase forms inside the low-density phase.
However, a completely analogous derivation works for a droplet of a low-density phase immersed 
in a high-density environment (e.g., vapor bubbles in water). 

\end{remark}

\begin{remark}
\label{rem2a}
Once we have derived the Gibbs-Thomson formula~\eqref{1.17}, we can also justify our choice of~$\pinfty$ for
the pressure ``above a planar interface.''
First let us note that, in~\eqref{1.17},~$\pinfty$ can be viewed as a convenient 
normalization constant---subtracting~\eqref{1.17}
for two different volumes, say~$V_1$ and~$V_2$, the quantity~$\pinfty$ completely factors out.
Moreover, if~$V_1\ll V_2$, the contribution of the droplet in~$V_2$ to such a difference will be negligible. 
Thus, in the limit when~$V_2\to\infty$ and~$V_1$ stays fixed, 
$p_{V_1}-p_{V_2}$ tends to~$p_{V_1}-\pinfty$ as expressed in~\eqref{1.17}.
Since also the droplet in~$V_2$ becomes more and more flat in this limit,~$\pinfty$ indeed represents the pressure ``above a planar~interface.''
\end{remark}

This concludes our heuristic derivation of the Gibbs-Thomson formula.
We reiterate that all of the above only makes good sense when~$\tV$ has been chosen such that
\begin{equation}
\label{tVagain}
V^{1-\frac2d+\frac 1{d+1}}\ll\tV\ll \deltaV\sim V^{1-\frac 1{d+1}}.
\end{equation} 
As is easily checked, these inequalities represent a non-trivial interval of values of~$\tV$.
In the next sections, where we will rigorously treat the case of the two-dimensional Ising lattice gas,
the inequality on the right-hand side will be guaranteed by taking~$\tV =\eta \deltaV$
and then performing the limits~$V\to\infty$ followed by~$\eta\to 0$.

\section{Rigorous results}
\label{sec2}\noindent
\vspace{-8mm}
\subsection{The model}
\label{sec2.1}\noindent
Throughout the remainder of this paper, we will focus our attention on the \emph{two-dimensional Ising lattice gas}. 
The latter refers to a system where each site of the square lattice~$\Z^2$ can be either vacant or occupied by one particle.
The state of each site is characterized by means of an occupation number~$n_x$ which is zero for a vacant site and
one for an occupied site. The formal Hamiltonian of the system can be written as
\begin{equation}
\label{H}
\mathscr{H}=-\sum_{\langle x, y \rangle}n_xn_y - \mu \sum_x n_x.
\end{equation}
Here~$\langle x,y\rangle$ denotes a nearest-neighbor pair on~$\Z^2$ and~$\mu$ plays the role of a chemical potential.
Note that the Hamiltonian describes particles with a hard-core repulsion and short-range attraction 
(with coupling constant set to unity).

The Gibbs measure (or Gibbs state) on particle configurations in a finite volume~$\Lambda\subset\Z^2$ is defined
using the finite-volume version of \eqref{H} and a boundary condition on the boundary of~$\Lambda$.
Explicitly, let~$\partial\Lambda$ be the set of sites in $\Z^2\setminus\Lambda$ that have a bond 
into $\Lambda$ and let $\mathscr{H}_\Lambda$  be the restriction of~$\mathscr{H}$ 
obtained by considering only pairs~$\{x,y\}\cap\Lambda\ne\emptyset$ in the first sum in~\eqref{H} 
and  sites~$x\in\Lambda$ in the second sum.
If~$n_\Lambda\in\{0,1\}^\Lambda$ is a configuration in~$\Lambda$ and~$n_{\partial\Lambda}$ is a boundary condition 
(i.e., a configuration on the boundary~$\partial\Lambda$ of~$\Lambda$), 
and if~$\mathscr{H}_\Lambda(n_\Lambda|n_{\partial\Lambda})$ is  the Hamiltonian for these two configurations, 
then the probability of~$n_\Lambda$ in the corresponding Gibbs measure is 
given by
\begin{equation}
\label{2.2}
P_\Lambda^{\,n_{\partial\Lambda},\beta,\mu}(n_\Lambda)
=\frac{e^{-\beta\mathscr{H}_\Lambda(n_\Lambda|n_{\partial\Lambda})}}
{\ZG^{\,n_{\partial\Lambda},\beta}(\mu,\Lambda)}.
\end{equation}
Here, as usual,~$\beta\ge0$ is the inverse temperature and the normalization constant, 
$\ZG^{\,n_{\partial\Lambda},\beta}(\mu,\Lambda)$, is the grand canonical partition function in~$\Lambda$ 
corresponding to the boundary condition~$n_{\partial\Lambda}$.
We recall that, according to the standard DLR-scheme~\cite{Georgii}, the system is at \emph{phase coexistence} 
if (depending on the boundary conditions and/or the sequence of volumes)
there is more than one infinite-volume limit of the measures in~\eqref{2.2}.
Of particular interest will be the measure in~$L\times L$ rectangular volume~$\Lambda_L\subset\Z^2$ and vacant (i.e.,~$n_{\partial\Lambda_L}\equiv0$) boundary condition. In this case we will denote the object from \eqref{2.2} by~$P_L^{\circ,\beta,\mu}$.

As is well known, the lattice gas model \eqref{H} is equivalent to the Ising magnet with the (formal) Hamiltonian
\begin{equation}
\label{HIs}
\mathscr{H}=-J\sum_{\langle x, y \rangle}\sigma_x\sigma_y - h\sum_x \sigma_x,
\end{equation}
coupling constant~$J=1/4$, external field~$h=\mu-2$ and the Ising spins ($\sigma_x$) related 
to the occupation variables ($n_x$) via~$\sigma_x=2n_x-1$.
The~$\pm$-symmetry of the Ising model also allows us to identify the regions of phase coexistence 
of the lattice gas model defined by \eqref{H}:
There is a value~$\betac=2\log (1+\sqrt 2) $
of the inverse temperature such that for~$\beta>\betac$
and $\mu=\mut=2$, there exist two distinct translation-invariant, extremal, ergodic, infinite-volume Gibbs states 
for the Hamiltonian~\eqref{H}---a ``liquid'' state characterized by an abundance of particles over vacancies 
and a ``gaseous'' state, characterized by an abundance of vacancies over occupied sites. In the Ising-spin language, 
these states correspond to the plus and minus states which in the lattice gas language translate to the states generated by the fully occupied or vacant boundary conditions. We will use~$\langle-\rangle_\beta^\circ$ 
and~$\langle-\rangle_\beta^\bullet$ to denote the expectation with respect to the (infinite-volume) ``gaseous'' 
and ``liquid'' state, respectively.

In order to discuss the Gibbs-Thomson formula in this model, we need to introduce the relevant quantities. 
Assuming~$\mu=\mut$ and~$\beta>\betac$, we will begin by defining the gas and liquid densities:
\begin{equation}
%\label{}
\rhoG=\rhoG(\beta)=\langle n_0\rangle_\beta^\circ
\quad\text{and}\quad
\rhoL=\rhoL(\beta)=\langle n_0\rangle_\beta^\bullet,
\end{equation}
where~$n_0$ refers to the occupation variable at the origin. Note that, by the plus-minus Ising symmetry, 
$\langle n_0\rangle_\beta^\circ=\langle1-n_0\rangle_\beta^\bullet$ and thus $\rhoL+\rhoG=1$. 
Next we will introduce the quantity~$\varkappa$ which is related to isothermal compressibility:
\begin{equation}
%\label{}
\varkappa=\sum_{x\in\Z^2}\bigl(\langle n_0n_x\rangle_\beta^\circ-\rhoG^2\bigr).
\end{equation}
The sum converges for all~$\beta>\betac$ by the exponential decay of truncated particle-particle correlations,
$|\langle n_xn_y\rangle_\beta^\circ-\rhoG^2|\le e^{-|x-y|/\xi}$, where~$\xi=\xi(\beta)<\infty$ denotes 
the correlation length.
The latter was proved in~\cite{CCS,Schonmann-Shlosman} in the context of the 2D~Ising model.

The last object we need to bring into play is the surface tension  or the interfacial free energy. 
In the 2D~Ising model, one can use several equivalent definitions. 
Since we will not need any of them explicitly, 
it suffices if we  just summarize the major concepts
as formulated,  more or less, in~\cite{DKS,Pfister}: 
First, for each~$\beta>\betac$, there is 
a continuous function~$\tau_\beta\colon\{\bn\in\R^2\colon|\bn|=1\}\to(0,\infty)$, 
called the \emph{microscopic surface tension}.
Roughly speaking, $\tau_\beta(\bn)$ is the cost per length of an interface with normal
vector~$\bn$ that separates a ``gaseous'' and ``liquid'' region.
This allows to introduce the so called {\em Wulff functional}~$\mathscr{W}_\beta$ that assigns 
to each rectifiable curve~$\varphi=(\varphi_t)$ in~$\R^2$ the value
\begin{equation}
\label{2.4}
\mathscr{W}_\beta(\varphi)=\int_\varphi \tau_\beta(\bn_t)\textd\bn_t.
\end{equation}
Here~$\bn_t$ is the normal vector to~$\varphi$ at the point~$\varphi_t$.

The quantity~$\mathscr{W}_\beta(\partial D)$ 
expresses the macroscopic cost of a droplet~$D$ with boundary~$\partial D$.  
Indeed, as has been established in the course of last few years~\cite{DKS,Pfister,Pf-Velenik,Ioffe1,Ioffe2,Bob+Tim}, 
the probability in the measure~$P_L^{\circ,\beta,\mut}$ that a droplet of ``liquid'' phase occurs whose shape is ``near'' that of the set~$D$ is given, to leading order, by 
$\exp\{-\mathscr{W}_\beta(\partial D)\}$.
Thus the ``most favorable'' droplet shape is obtained
by minimizing~$\mathscr{W}_\beta(\partial D)$ over all~$D$ with a given volume. 
Using~$W$ to denote the minimizing set with a \emph{unit volume}
(which can be explicitly constructed~\cite{Wulff,Curie,Gibbs}), 
we~define
\begin{equation}
\label{2.5}
w_1(\beta)=\mathscr{W}_\beta(\partial W).
\end{equation}
By well-known properties of the surface tension, we have~$w_1(\beta)>0$ once~$\beta>\betac$.
We note that, as in the heuristic section---see Remark~\ref{rem1}---the customary factor~$1/\beta$  
is incorporated into~$\tau_\beta$ in our definition of the surface tension.

\begin{remark}
\label{rem3}
For those more familiar with the magnetic terminology, let us pause 
to identify the various quantities in Ising language:
First, if~$\mstar(\beta)$ is the  \emph{spontaneous magnetization}, 
then we have~$\rhoG(\beta)=\frac12(1-\mstar(\beta/4))$ 
and~$\rhoL(\beta)=\frac12(1+\mstar(\beta/4))$. Similarly, if~$\chi(\beta)$ denotes 
the \emph{magnetic susceptibility} in the Ising spin system, then~$\varkappa(\beta)=\varkappa(\beta/4)/4$. 
Finally, the quantity~$w_1(\beta)$ corresponds exactly to the similar quantity for the spin system at a quarter 
of the inverse temperature.
\end{remark}

\subsection{Known facts}
\label{sec2.2}\noindent
Here we will review some of the rigorous results concerning the 2D~Ising lattice gas in a finite volume 
and a fixed number of particles. In the language of statistical mechanics, this corresponds to the \emph{canonical} 
ensemble. The stated theorems are transcribes of the corresponding results from~\cite{bigBCK}.

\smallskip
Recall our notation~$P_L^{\circ,\beta,\mu}$ for the Gibbs state in~$L\times L$ rectangular box~$\Lambda_L$ and vacant boundary conditions on~$\partial\Lambda_L$.
Let~$(v_L)$ be a sequence of positive numbers tending to infinity in such a way that~$v_L^{3/2}/|\Lambda_L|$ tends
to a finite non-zero limit. 
In addition,  suppose that~$(v_L)$ is such that~$\rhoG|\Lambda_L|+(\rhoL-\rhoG)v_L$ is a number from 
$\{0,1,\dots,|\Lambda_L|\}$ for all~$L$.
For any configuration~$(n_x)$ in~$\Lambda_L$, let~$N_L$ denote the total number of particles in~$\Lambda_L$,~i.e.,
\begin{equation}
\label{2.6}
N_L=\sum_{x\in\Lambda_L}n_x.
\end{equation}
Our first theorem concerns the large-deviation asymptotic for the random variable~$N_L$. 
The following is a rigorous version of the claim~\eqref{PV1}, which, more or less, 
is Theorem~1.1 from~\cite{bigBCK}.

\renewcommand{\thetheorem}{\Alph{theorem}}

\begin{theorem}
\label{thmA} 
Let~$\beta>\betac$ and let the sequence~$(v_L)$ and 
the quantities~$\rhoG=\rhoG(\beta)$, 
$\rhoL=\rhoL(\beta)$,~$\varkappa=\varkappa(\beta)$, and~$w_1=w_1(\beta)$ be as defined previously.
Suppose that  the limit
\begin{equation}
\label{Delta-lim} 
\Delta=\frac{(\rhoL-\rhoG)^2}{2\varkappa w_1}
\,\lim_{L\to\infty}\frac{\,v_L^{3/2}}{|\Lambda_L|}
\end{equation} 
exists with~$\Delta\in(0,\infty)$. Then
\begin{equation}
\label{LDP} 
\lim_{L\to\infty}\frac1{\sqrt{v_L}}\,\log
P_L^{\circ,\beta,\mut}\bigl(N_L=\rhoG|\Lambda_L|+(\rhoL-\rhoG)v_L\bigr)=
-w_1\inf_{0\le \lambda\le1}\Phi_\Delta(\lambda),
\end{equation}
where~$\Phi_\Delta(\lambda)=\sqrt\lambda+\Delta(1-\lambda)^2$.
\end{theorem}

We proceed by a description of the typical configurations in the conditional measure
\begin{equation}
\label{PLcond}
P_L^{\circ,\beta,\mut}\bigl(\cdot\big|N_L=\rhoG|\Lambda_L|+(\rhoL-\rhoG)v_L\bigr),
\end{equation} 
which, we note, actually does not depend on the choice of the chemical potential.
Our characterization will be based on the notion of \emph{Peierls' contours}:
Given a particle configuration, let us place a dual bond in the middle of each direct bond connecting
an occupied and a vacant site.
These dual bonds can be connected into self-avoiding polygons by applying an appropriate ``rounding rule,'' 
as discussed in~\cite{DKS} and illustrated in, e.g.,~Fig.~1 of~\cite{bigBCK}.
Given a contour~$\gamma$, let~$V(\gamma)$ denote the set of sites enclosed by~$\gamma$. 
In accord with~\cite{bigBCK}, we also let~$\diam\gamma$ denote the diameter of the set~$V(\gamma)$
in the~$\ell_2$ metric on~$\Z^2$. If~$\Gamma$ is a collection of contours, we say that~$\gamma\in\Gamma$ 
is an \emph{external} contour if it is not surrounded by any other contour from~$\Gamma$.

While ``small'' contours are just natural fluctuations within a given phase,
``large'' contours should somehow be interpreted as droplets. It turns out that the corresponding scales 
are clearly separated with no intermediate contours present in typical configurations.
The following is essentially the content of Theorem~1.2 and Corollary~1.3 from~\cite{bigBCK}.

\begin{theorem}
\label{thmB}
Let~$\beta>\betac$ and let the sequence~$(v_L)$ and the quantities~$\rhoG=\rhoG(\beta)$, 
$\rhoL=\rhoL(\beta)$,~$\varkappa=\varkappa(\beta)$, and~$w_1=w_1(\beta)$ be as defined previously. 
Suppose that the limit in~\eqref{Delta-lim} exists with~$\Delta\in(0,\infty)$ and let~$\Deltac=\frac12(3/2)^{3/2}$. 
There exists a number~$K=K(\beta,\Delta)<\infty$ such that, for each~$\epsilon>0$ and~$L\to\infty$, 
the following holds with probability tending to one in the distribution
\eqref{PLcond}:

\smallskip\noindent
(1) If~$\Delta<\Deltac$, then all contours~$\gamma$ satisfy~$\diam\gamma\le K\log L$.

\smallskip\noindent
(2) If~$\Delta>\Deltac$, then there exists a unique contour~$\gamma_0$ with
\begin{equation}
\label{2.9a}
\lambda_\Delta v_L(1-\epsilon)\le\bigl|V(\gamma_0)\bigr|
\le\lambda_\Delta v_L(1+\epsilon)
\end{equation}
and
\begin{equation}
\label{2.9}
\rhoL\lambda_\Delta v_L(1-\epsilon)\le\sum_{x\in V(\gamma_0)}n_x
\le\rhoL\lambda_\Delta v_L(1+\epsilon),
\end{equation}
where~$\lambda_\Delta$ is the largest solution to the equation
\begin{equation}
\label{lambeq}
4\Delta\sqrt\lambda(1-\lambda)=1
\end{equation}
in~$[0,1]$. 
Moreover, all the other external contours~$\gamma\ne\gamma_0$ satisfy~$\diam\gamma\le K\log L$.
\end{theorem}

\renewcommand{\thetheorem}{\arabic{section}.\arabic{theorem}}
\setcounter{theorem}{0}

\begin{remark}
\label{rem4}
We note that, in the case~$\Delta=\Deltac$, there is at most one large external contour satisfying the bounds
\twoeqref{2.9a}{2.9}, or no contour beyond~$K\log L$ at all.
The details of what exactly happens when $\Delta=\Deltac$ have not, at present, been quantified---presumably, these will depend on the asymptotic of the sequence~$v_L$.

\end{remark}

\begin{remark}
\label{rem4a}
One additional piece of information we could add about the contour~$\gamma_0$ is that its
macroscopic shape asymptotically optimizes the Wulff functional, see~\twoeqref{2.4}{2.5}.
While the shape of the unique large contour plays no essential role in this paper
(it appears  implicitly in the value $w_1$) 
we note that statements of this sort were the basis of the (microscopic) \emph{Wulff construction}, 
initiated in~\cite{ACC,DKS} for the case of 2D~Ising model and percolation.
These~2D results were later extended in~\cite{Pfister,Pf-Velenik,DS,Ioffe1,Ioffe2,Bob+Tim}.
The techniques developed in these papers have been instrumental for the results of~\cite{bigBCK}, 
which addresses the regime that is ``critical'' for droplet formation.
Recently, extensions going beyond two spatial dimensions have also been 
accomplished~\cite{Cerf,Bodineau,Cerf-Pisztora}.
We refer to~\cite{BIV} and~\cite{bigBCK} for more information on  the subject.
\end{remark}

\subsection{Gibbs-Thomson formula(s) for 2D~Ising lattice gas}
\label{sec2.3}\noindent
Now we are finally in a position to state our rigorous version of the Gibbs-Thomson formula for the 2D~Ising
lattice gas. We will begin with the formula for the difference of the densities, which is, more or less, an 
immediate corollary of Theorem~\ref{thmB}.

\begin{theorem}
\label{thm1}
Let~$\beta>\betac$ and let the sequence~$(v_L)$ and the  
quantities $\rhoG=\rhoG(\beta)$,
$\rhoL=\rhoL(\beta)$, $\varkappa=\varkappa(\beta)$,  
and $w_1=w_1(\beta)$ be as defined previously.
Let~$\Delta\in(0,\infty)$ be as in~\eqref{Delta-lim}.
Suppose that~$\Delta>\Deltac=\frac12(3/2)^{3/2}$ and  
let~$\lambda_\Delta$ be the largest solution
of the equation~\eqref{lambeq} in
the interval~$[0,1]$.
Let $\AA_{\epsilon,L}$ be the set of configurations  
$(n_x)_{x\in\Lambda_L}$ that contain a unique  
large external contour~$\gamma_0$---as described in  
Theorem~\ref{thmB}---obeying \twoeqref{2.9a}{2.9}, and whose particle density  
in the exterior of~$\gamma_0$,
\begin{equation}
%\label{}
\rho_\ext(\gamma_0)=\frac1{|\Lambda_L\setminus V(\gamma_0)|}
\sum_{x\in \Lambda_L\smallsetminus V(\gamma_0)}n_x,
\end{equation}
satisfies the bounds
\begin{equation}
\label{2.11}
\frac12\,\frac{\varkappa w_1}{\rhoL-\rhoG}
\frac1{|V(\gamma_0)|^{1/2}}(1-\epsilon)\le
\rho_\ext(\gamma_0)
-\rhoG
\le\frac12\,\frac{\varkappa  
w_1}{\rhoL-\rhoG}\frac1{|V(\gamma_0)|^{1/2}}(1+\epsilon).
\end{equation}
Then, for each~$\epsilon>0$, we have
\begin{equation}
\lim_{L\to\infty}P_L^{\circ,\beta,\mut}(\AA_{\epsilon,L}|N_L=\rhoG|\Lambda_L| 
+(\rhoL-\rhoG)v_L)=1.
\end{equation}
\end{theorem}

\smallskip
\begin{remark}
\label{rem5}
We note that, up to the~$\epsilon$ corrections,~\eqref{2.11} is exactly~\eqref{barrhoB} for~$d=2$.
Indeed, by Theorem~\ref{thmB} we know that~$|V(\gamma_0)|=\lambda_\Delta v_L(1+o(1))$ and the two formulas
are identified by noting that~$\deltaV$ corresponds to~$v_L$ in our setting.
Due to the underlying lattice, the Wulff droplet is undoubtedly not circular for any $\beta>\betac$ 
and the better-known form~\eqref{barrhoC} of the (density) Gibbs-Thomson formula does not apply.
\end{remark}

In order to state our version of the Gibbs-Thomson formula for the pressure, we will first need to define the
pressure ``above a curved interface''---not to mention the planar interface. We will closely follow 
the heuristic definitions \twoeqref{pV}{pinf}.
Let us consider a sequence~$(\Lambda_L')$ of squares in~$\Z^2$ satisfying
\begin{equation}
\label{TLa}
\Lambda_L'\supset\Lambda_L
\quad\text{but}\quad
\Lambda_L'\ne\Lambda_L
\end{equation}
for all~$L$.
Let~$\ZC^{\circ,\beta}(N,\Lambda)$ denote the \emph{canonical} partition function in~$\Lambda$ with~$N$
particles, inverse temperature~$\beta$ and the vacant boundary condition. This quantity is computed 
by summing the Boltzmann factor,  
\begin{equation}
%\label{}
\exp\biggl\{\mkern6mu\beta\sum_{\begin{subarray}{c}
\langle x,y\rangle\\x,y\in\Lambda
\end{subarray}}n_xn_y
\biggr\},
\end{equation}
over all configurations~$(n_x)$ with~$\sum_{x\in\Lambda}n_x=N$.
Then we let
\begin{equation}
\label{2.17b}
\pL=\frac1\beta\frac1{|\Lambda_L'\setminus\Lambda_L|}
\log\frac{\ZC^{\circ,\beta}(\rhoG|\Lambda_L|+(\rhoL-\rhoG)v_L,\Lambda_L')}
{\ZC^{\circ,\beta}(\rhoG|\Lambda_L|+(\rhoL-\rhoG)v_L,\Lambda_L)}.
\end{equation}
As in the heuristic section, the quantity~$\pL$ depends on the sequences~$(\Lambda_L')$,~$(v_L)$, 
inverse temperature~$\beta$, and also the boundary condition---all of which is notationally suppressed.

For the pressure ``above a planar interface,'' again we will simply use the pressure of the pure (gaseous) phase.
If~$\Lambda\subset\Z^2$ is a finite set, we let~$\ZG^{\circ,\beta}(\mu,\Lambda)$ denote the \emph{grand canonical} partition function
in~$\Lambda$ corresponding to the chemical potential~$\mu$ and vacant boundary condition.
Recalling that $\mut=2$, we~define
\begin{equation}
\label{2.16}
\pinfty = \frac1\beta\lim_{L\to\infty}\frac1{|\Lambda_L|}\log 
\ZG^{\circ,\beta}(\mut,\Lambda_L),
\end{equation}
where the limit exists by standard subadditivity arguments. 

Suppose that~$\Delta>\Deltac$ and let us consider the event~$\BB_{\epsilon,L}$ collecting all configurations in~$\Lambda_L$ that have a unique ``large'' contour~$\gamma_0$, as described in Theorem~\ref{thmB}, 
such that, in addition to~\twoeqref{2.9a}{2.9}, the  volume~$V(\gamma_0)$ satisfies the inequalities
\begin{equation}
\label{2.17}
\frac12\,\frac{\rhoG w_1}{\rhoL-\rhoG}\frac1{|V(\gamma_0)|^{1/2}}(1-\epsilon)\le
\beta(\pL-\pinfty)
\le\frac12\,\frac{\rhoG w_1}{\rhoL-\rhoG}\frac1{|V(\gamma_0)|^{1/2}}(1+\epsilon).
\end{equation}
Somewhat informally, the event~$\BB_{\epsilon,L}$ represents the configurations for which 
the Gibbs-Thomson formula for pressure holds up to an~$\epsilon$ error.
The next theorem shows that, as $L\to\infty$, these configurations exhaust all of the conditional measure~\eqref{PLcond}:

\begin{theorem}
\label{thm2}
Let~$\beta>\betac$ and let the sequence~$(v_L)$ and the quantities~$\rhoG=\rhoG(\beta)$, 
$\rhoL=\rhoL(\beta)$,~$\varkappa=\varkappa(\beta)$, and~$w_1=w_1(\beta)$ be as defined previously.
Let~$\Delta\in(0,\infty)$ be as in~\eqref{Delta-lim}.
Suppose that~$\Delta>\Deltac=\frac12(3/2)^{3/2}$ and let~$\lambda_\Delta$ be 
the largest solution to~\eqref{lambeq} in~$[0,1]$. 
For each~$\epsilon>0$, there exists a number~$\eta_0>0$ such that if~$(\Lambda_L')$ 
is a sequence of squares in~$\Z^2$ satisfying~\eqref{TLa} and
\begin{equation}
\label{TL}
\lim_{L\to\infty}\frac{|\partial\Lambda_L'|-|\partial\Lambda_L|}{|\Lambda_L'\setminus\Lambda_L|}
\sqrt{v_L}=0
\quad\text{and}\quad
\lim_{L\to\infty}\frac{|\Lambda_L'\setminus\Lambda_L|}{v_L}=\eta\in(0,\eta_0],
\end{equation}
then
\begin{equation}
\label{2.19}
\lim_{L\to\infty}P_L^{\circ,\beta,\mut}\bigl(\BB_{\epsilon,L}|N_L=\rhoG|\Lambda_L|+(\rhoL-\rhoG)v_L\bigr)=1.
\end{equation}
\end{theorem}

\begin{remark}
\label{rem6}
As before, since~$|V(\gamma_0)|=\lambda_\Delta v_L(1+o(1))$, the equality \eqref{2.19} 
is a rigorous version of~\eqref{1.17}  for the case at hand. 
The rate at which the limit in~\eqref{2.19} is achieved depends---among other things---on the rate 
of the convergence in~\eqref{TL}.
We note that the constraints~\eqref{TL} correspond to the bounds in~\eqref{tVagain}.
In particular, there is a non-trivial set of sequences~$(\Lambda_L')$ for which both limits in~\eqref{TL} 
are exactly as prescribed.
Finally, the restriction that~$\eta>0$ in~\eqref{TL} is due to the fact that from \cite{bigBCK} we have essentially 
no control on the rate of convergence in~\eqref{LDP}.
Thus, to allow the second limit in~\eqref{TL} to be zero, we would have to do a little extra work 
in order to clarify the rate at which the limits in~\eqref{TL} and~\eqref{LDP} are achieved.
\end{remark}
\vspace{-2mm}

\section{Proofs of main results}
\label{sec3}
\subsection{Proofs of Theorems~\ref{thm1} and~\ref{thm2}}
\label{sec3.1}\noindent
In this section we provide the proofs of our main results. We will commence with Theorem~\ref{thm1}:

\begin{proofsect}{Proof of Theorem~\ref{thm1}}
The proof closely follows the heuristic calculation from Section~\ref{sec1.2}.
Fix an~$\epsilon>0$ and let us restrict our attention to particle configurations containing 
a unique external contour $\gamma_0$ and satisfying the bounds \twoeqref{2.9a}{2.9}.
Recall the definition~\eqref{2.6} of the quantity~$N_L$.
We will show that, under the condition
\begin{equation}
\label{podminka}
N_L=\rhoG|\Lambda_L|+(\rhoL-\rhoG)v_L,
\end{equation}
any such configuration is, for a suitable $\epsilon'>0$, contained in
$\AA_{\epsilon',L}$ for all~$L$.
Let
\begin{equation}
%\label{}
N_\ext(\gamma_0)=\sum_{x\in\Lambda_L\smallsetminus V(\gamma_0)}n_x.
\end{equation}
The inequalities in~\eqref{2.9} then directly imply
\begin{equation}
%\label{}
\bigl| N_\ext(\gamma_0)-(N_L-\rhoL\lambda_\Delta v_L)\bigr|\le \epsilon\rhoL\lambda_\Delta v_L.
\end{equation}
Since we work with a measure conditioned on the event \eqref{podminka}, we can~write
\begin{equation}
%\label{}
N_L-\rhoL\lambda_\Delta v_L=
\rhoG\bigl(|\Lambda_L|-\lambda_\Delta v_L\bigr)+(\rhoL-\rhoG)(1-\lambda_\Delta)v_L.
\end{equation}
But~$|\Lambda_L|-\lambda_\Delta v_L=|\Lambda_L\setminus V(\gamma_0)|+(|V(\gamma_0)|-\lambda_\Delta v_L)$
and by~\eqref{2.9a}, the second term is no larger than~$\epsilon\lambda_\Delta v_L$.
Combining the previous estimates, we derive the bound
\begin{equation}
\label{3.3}
\bigl| N_\ext(\gamma_0)-\rhoG|\Lambda_L\setminus V(\gamma_0)|
-(\rhoL-\rhoG)(1-\lambda_\Delta)v_L\bigr|
\le \epsilon\lambda_\Delta v_L,
\end{equation}
where we also used (inessentially) that~$\rhoL+\rhoG=1$ (and thus $\rho_g\le 1$).

The first two terms in the absolute value on the left-hand side represent the difference 
between~$\rho_\ext(\gamma_0)$ and~$\rhoG$ while the third term is exactly the Gibbs-Thomson correction. 
Indeed, dividing \eqref{3.3} by $|\Lambda_L\setminus V(\gamma_0)|$ and noting that, by definition, 
$N_\ext(\gamma_0)=\rho_\ext(\gamma_0)|\Lambda_L\setminus V(\gamma_0)|$,~we~get
\begin{equation}
\label{3.3a}
\biggl| \rho_\ext(\gamma_0)-\rhoG
-(\rhoL-\rhoG)\frac{(1-\lambda_\Delta)v_L}{|\Lambda_L\setminus V(\gamma_0)|}\biggr|
\le \frac{\epsilon\lambda_\Delta v_L}{|\Lambda_L\setminus V(\gamma_0)|}.
\end{equation}
Since both the Gibbs-Thomson correction---which arises from the last term 
in the above absolute value---and the error term on the right-hand side are proportional 
to~$v_L/|\Lambda_L\setminus V(\gamma_0)|$, the desired bound~\eqref{2.11} will follow 
with \emph{some} $\epsilon>0$ once we show that
\begin{equation}
\label{3.4}
(\rhoL-\rhoG)\frac{(1-\lambda_\Delta)v_L}{|\Lambda_L\setminus V(\gamma_0)|}
=\frac12\,\frac{\varkappa w_1}{\rhoL-\rhoG}\frac1{\sqrt{\lambda_\Delta v_L}}\bigl(1+o(1)\bigr),
\quad L\to\infty.
\end{equation}
To prove~\eqref{3.4}, we note that~$|\Lambda_L\setminus V(\gamma_0)|/|\Lambda_L|=1+o(1)$,
which using~\eqref{Delta-lim} allows us to write 
\begin{equation}
%\label{}
\frac{v_L}{|\Lambda_L\setminus V(\gamma_0)|}
=\frac{2\varkappa w_1}{(\rhoL-\rhoG)^{2}}\frac\Delta{\sqrt{v_L}}\bigl(1+o(1)\bigr),
\quad L\to\infty.
\end{equation}
Using~\eqref{lambeq} in the form~$\Delta(1-\lambda_\Delta)=1/(4\sqrt{\lambda_\Delta})$,
we get rid of the factor of~$\Delta$, whereby \eqref{3.4} follows.
Since the $o(1)$~term in \eqref{3.4} is uniformly small for all configurations satisfying \twoeqref{2.9a}{2.9}, 
the bounds \eqref{2.11} hold once~$L$ is sufficiently large.
\end{proofsect}

In order to prove our Gibbs-Thomson formula for the pressure, we will need the following representation of
the grand canonical partition function:

\begin{theorem}
\label{thm3.1}
Let~$\beta>\betac$ and let~$\pinfty$ be as in~\eqref{2.16}. 
There exists a number~$\tau_{\text{\rm wall}}^\circ\in\R$ and, for each~$\theta\in(1,\infty)$, 
also a constant~$C(\beta,\theta)<\infty$ such that 
\begin{equation}
\label{eq3.9}
\bigl|\log\ZG^{\circ,\beta}(\mut,\Lambda)-\beta \pinfty|\Lambda|
-\tau_{\text{\rm wall}}^\circ|\partial\Lambda|\bigr|\le C(\beta,\theta)
\end{equation}
holds for all rectangular volumes~$\Lambda\subset\Z^2$ whose
aspect ratio lies in the interval~$(\theta^{-1},\theta)$.
\end{theorem}

Clearly, Theorem~\ref{thm3.1} is a rigorous version of the formula~\eqref{ZGrep}.
Such things are well known in the context of low-temperature expansions, see, e.g.,~\cite{Borgs-Kotecky}.
Here we are using  expansion techniques in conjunction with correlation inequalities 
to get the claim  ``down to~$\betac$.''
However, the full argument would detract from the main line of thought, so the proof is postponed 
to Section~\ref{sec3.2}.

\begin{proofsect}{Proof of Theorem~\ref{thm2}}
We will again closely follow the heuristic derivation from Section~\ref{sec1.2}.
First we note that, using Theorem~\ref{thm3.1}, we have
\begin{equation}
%\label{}
\biggl|\beta \pinfty - \frac1{|\Lambda_L'\setminus\Lambda_L|}
\log\frac{\ZG^{\circ,\beta}(\mut,\Lambda_L')}{\ZG^{\circ,\beta}(\mut,\Lambda_L)}\biggr|
\le |\tau_{\text{\rm wall}}^\circ|
\frac{|\partial\Lambda_L'|-|\partial\Lambda_L|}{|\Lambda_L'\setminus\Lambda_L|}
+\frac{2C(\beta,\theta)}{|\Lambda_L'\setminus\Lambda_L|}.
\end{equation}
Introducing the shorthand
\begin{equation}
%\label{}
\BbbP_\Lambda(N)=P_\Lambda^{\circ,\beta,\mut}\Bigl(\,\sum_{x\in\Lambda}n_x=N\Bigr),
\end{equation}
invoking the assumption on the left of~\eqref{TL} and applying~\eqref{2.17b}, this allows us to write
\begin{equation}
\label{3.10}
\beta(\pL-\pinfty)=\frac1{|\Lambda_L'\setminus\Lambda_L|}
\log\frac
{\BbbP_{\Lambda_L'}(\rhoG|\Lambda_L|+(\rhoL-\rhoG)v_L)}
{\BbbP_{\Lambda_L}(\rhoG|\Lambda_L|+(\rhoL-\rhoG)v_L)}+o(v_L^{-1/2}),
\quad L\to\infty.
\end{equation}
Now, by Theorem~\ref{thmA} we have
\begin{equation}
\label{pD1}
\log\BbbP_{\Lambda_L}\bigl(\rhoG|\Lambda_L|+(\rhoL-\rhoG)v_L\bigr)
=-w_1\bigl(\Phi_\Delta^\star+o(1)\bigr)\sqrt{v_L},
\quad L\to\infty,
\end{equation} 
where~$\Phi_\Delta^\star$ is the absolute minimum of~$\Phi_\Delta(\lambda)$ for~$\lambda\in[0,1]$. 
As to the corresponding probability for~$\Lambda_L'$, we first note that
\begin{equation}
%\label{}
\rhoG|\Lambda_L|+(\rhoL-\rhoG)v_L=\rhoG|\Lambda_L'|+(\rhoL-\rhoG)\alpha_Lv_L,
\end{equation}
where
\begin{equation}
%\label{}
\alpha_L=1-\frac{\rhoG}{\rhoL-\rhoG}\frac{|\Lambda_L'\setminus\Lambda_L|}{v_L}.
\end{equation}
By our assumption on the right-hand side of~\eqref{TL}, 
$\alpha_L$ converges to a number~$\alpha$ given by~$\alpha=1-\frac{\rhoG}{\rhoL-\rhoG}\eta$.
Again using Theorem~\ref{thmA}, we can write
\begin{equation}
\label{pD2}
\log\BbbP_{\Lambda_L}\bigl(\rhoG|\Lambda_L|+(\rhoL-\rhoG)v_L\bigr)
=-w_1\bigl(\Phi_{\alpha^{3/2}\Delta}^\star
+o(1)\bigr)\sqrt\alpha\sqrt{v_L},\quad L\to\infty.
\end{equation} 
A simple calculation---of the kind leading to~\eqref{1.15}---now shows that
\begin{equation}
%\label{}
\sqrt\alpha\Phi_{\alpha^{3/2}\Delta}^\star-\Phi_\Delta^\star=
\frac\eta2\frac{\rhoG}{\rhoL-\rhoG}\frac1{\sqrt{\lambda_\Delta}}+O(\eta^2),
\qquad \eta\downarrow0,
\end{equation}
while~\eqref{TL} implies that
\begin{equation}
%\label{}
\frac{\sqrt{v_L}}{|\Lambda_L'\setminus\Lambda_L|}=\frac1{\sqrt{v_L}}\frac1\eta\bigl(1+o(1)\bigr),
\qquad L\to\infty.
\end{equation}
Plugging these equations, along with~\eqref{pD1} and~\eqref{pD2}, into~\eqref{3.10}, we have
\begin{equation}
\label{3.17}
\beta(\pL-\pinfty)=\frac12\frac{\rhoG w_1}{\rhoL-\rhoG}\frac1{\sqrt{\lambda_\Delta v_L}}
\Bigl(1+\frac{o(1)}\eta+O(\eta)\Bigr),
\end{equation}
where~$o(1)$ denotes a quantity tending to zero as~$L\to\infty$ while~$O(\eta)$ is a quantity independent of~$L$ 
and tending to zero at least as fast as~$\eta$ in the limit~$\eta\downarrow0$.
Equation~\eqref{3.17} shows that, once~$L$ is sufficiently large, 
a particle configuration satisfying the bounds~\eqref{2.9a} from Theorem~\ref{thmB}
will also satisfy the bounds~\eqref{2.17}. 
The limit~\eqref{2.19} is then a simple conclusion of Theorem~\ref{thmB}.
\end{proofsect}

\subsection{Representation of the partition function}
\label{sec3.2}\noindent
The goal of this section is to prove Theorem~\ref{thm3.1}.
As already mentioned, we will employ two basic techniques: cluster expansion and correlation inequalities.
The basic strategy of the proof is as follows. First we pick a large negative number~$\mu_0<\mut$ 
and use cluster expansion to establish a corresponding representation for 
the partition function~$\ZG^{\circ,\beta}(\mu_0,\Lambda_L)$. 
Then, as a second step, we invoke correlation inequalities to prove a similar representation for the ratio of the
partition functions~$\ZG^{\circ,\beta}(\mu_0,\Lambda_L)$ and~$\ZG^{\circ,\beta}(\mut,\Lambda_L)$.
Essential for the second step will be the GHS inequality and the exponential decay of correlations for all~$\beta>\betac$.
Combining these two steps, the desired representation will be proved.

\smallskip
Let~$\pinfty(\mu)$ denote the pressure corresponding to the chemical potential~$\mu$, which is defined 
by the limit as in \eqref{2.16} where~$\mut$ is replaced by~$\mu$. 
(Throughout this derivation, we will keep~$\beta$ fixed and suppress it notationally whenever possible.)
The first step in the above strategy can then be formulated as follows:

\begin{lemma}
\label{lemma3.2}
Let~$\beta>\betac$ and let~$\pinfty(\mu)$ be as defined above. 
For each~$\theta\in(1,\infty)$ and 
each sufficiently large negative~$\mu_0$,  there exists a number 
$\tau_1^\circ(\mu_0)\in\R$ 
and a constant $C_1(\beta,\mu_0,\theta)<\infty$ such that
\begin{equation}
\label{Q1}
\bigl|\log\ZG^{\circ,\beta}(\mu_0,\Lambda)-\beta \pinfty(\mu_0)|\Lambda|
-\tau_1^\circ(\mu_0)|\partial\Lambda|\bigr|\le C_1(\beta,\mu_0,\theta)
\end{equation}
holds for each rectangular volume~$\Lambda\subset\Z^2$ whose aspect ratio lies in the interval~$(\theta^{-1},\theta)$.
\end{lemma}

To implement the second step of the proof, we need to study the ratio of the partition functions 
with chemical potentials~$\mut$ and~$\mu_0$.
Let~$\Lambda$ be a finite rectangular volume in~$\Z^2$ and let~$\langle-\rangle_\Lambda^{\circ,\beta,\mu}$ 
denote the expectation with respect to the measure in~\eqref{2.2} with vacant boundary condition. 
Let~$N_\Lambda=\sum_{x\in\Lambda}n_x$.
For any~$\mu_0<\mut$ we then have
\begin{equation}
\label{logZZ}
\log\frac{\ZG^{\circ,\beta}(\mut,\Lambda_L)}{\ZG^{\circ,\beta}(\mu_0,\Lambda_L)}
=\int_{\mu_0}^{\mut}\langle N_\Lambda\rangle_\Lambda^{\circ,\beta,\mu}\textd\mu
\end{equation}
and
\begin{equation}
\label{pp}
\beta\bigl(\pinfty(\mut)-\pinfty(\mu_0)\bigr)=\int_{\mu_0}^{\mut}\langle n_0\rangle^{\circ,\beta,\mu}\textd\mu.
\end{equation}
where~$\langle-\rangle^{\circ,\beta,\mu}$ denotes the infinite-volume limit (which we are assured exists) 
of the state~$\langle-\rangle_\Lambda^{\circ,\beta,\mu}$.
(Note that \eqref{pp} is true with \emph{any} infinite-volume Gibbs state substituted.)
Combining \twoeqref{logZZ}{pp}, we thus get
\begin{equation}
\label{3.19}
\log\frac{\ZG^{\circ,\beta}(\mut,\Lambda_L)e^{-\beta \pinfty(\mut)|\Lambda|}}
{\ZG^{\circ,\beta}(\mu_0,\Lambda_L)e^{-\beta \pinfty(\mu_0)|\Lambda|}}
=\int_{\mu_0}^{\mut}\bigl(\langle N_\Lambda\rangle_\Lambda^{\circ,\beta,\mu}
-|\Lambda|\langle n_0\rangle^{\circ,\beta,\mu}\bigr)\textd\mu.
\end{equation}
To derive the desired representation, we need to show that the integrand is proportional to~$|\partial\Lambda|$, 
up to an error which does not depend on~$\Lambda$.
This estimate is provided in the following lemma:

\begin{lemma}
\label{lemma3.3}
Let~$\beta>\betac$ and~$\theta\in(1,\infty)$.
There exists a constant~$C_2(\beta,\theta)<\infty$ and a bounded function~$\tau_2^\circ\colon(-\infty,\mut]\to[0,\infty)$ 
such that
\begin{equation}
\label{Q2}
\bigl|\langle N_\Lambda\rangle_\Lambda^{\circ,\beta,\mu}
-|\Lambda|\langle n_0\rangle^{\circ,\beta,\mu}-|\partial\Lambda|\tau_2^\circ(\mu)\bigr|\le C_2(\beta,\theta),
\qquad\mu\in(-\infty,\mut],
\end{equation}
holds for each rectangular volume~$\Lambda\subset\Z^2$ whose aspect ratio lies in the interval~$(\theta^{-1},\theta)$.
\end{lemma}

\smallskip
Lemma~\ref{lemma3.2} will be proved in Section~\ref{sec3.3} and Lemma~\ref{lemma3.3} in Section~\ref{sec3.4}.
With the two lemmas in the hand, the proof of Theorem~\ref{thm3.1} is easily concluded:

\begin{proofsect}{Proof of Theorem~\ref{thm3.1}}
Let~$\theta\in(1,\infty)$ and let~$\Lambda$ be a rectangular volume whose aspect ratio lies 
in the interval~$(\theta^{-1},\theta)$.
Fix~$\mu_0$ to be so large 
(and negative) that Lemma~\ref{lemma3.2} holds and let~$Q_1(\mu_0)$ denote the quantity 
in the absolute value in~\eqref{Q1}.
For each~$\mu\in[\mu_0,\mut]$, let~$Q_2(\mu)$ denote the quantity inside the absolute value in~\eqref{Q2}. 
Let us define
\begin{equation}
%\label{}
\tau_{\text{\rm wall}}^\circ=\tau_1^\circ(\mu_0)+\int_{\mu_0}^{\mut}\tau_2^\circ(\mu)\textd\mu.
\end{equation}
A simple calculation combining~\eqref{Q1},~\eqref{Q2} with~\eqref{3.19} then shows that
\begin{equation}
%\label{}
\log\ZG^{\circ,\beta}(\mu,\Lambda)-\beta \pinfty(\mut)|\Lambda|-\tau_{\text{\rm wall}}^\circ|\partial\Lambda|
=Q_1(\mu_0)+\int_{\mu_0}^{\mut}Q_2(\mu)\textd\mu.
\end{equation}
Using~\eqref{Q1} and~\eqref{Q2}, we easily establish that the absolute value of the quantity on right-hand side is no
larger than~$C(\beta,\theta)=C_1(\beta,\mu_0,\theta)+(\mut-\mu_0)C_2(\beta,\theta)$.
\end{proofsect}

\subsection{Cluster expansion}
\label{sec3.3}\noindent
Here we will rewrite the grand canonical partition function in terms of a polymer model, then we will collect a few facts from the theory of cluster expansions and assemble them into the proof of Lemma~\ref{lemma3.2}.
The substance of this section is very standard---mostly siphoned from \cite{Kotecky-Preiss}---so 
the uninterested reader may wish to consider skipping the entire section on a first reading.

\smallskip
We begin by defining the polymer model.
Given a configuration~$n_\Lambda$ in~$\Lambda$, let us call two distinct sites of~$\Z^2$ connected 
if they are nearest-neighbors and are both occupied in the configuration~$n_\Lambda$.
A \emph{polymer} is then defined as a connected component of occupied sites.
Two polymers are called \emph{compatible} if their union is not connected.
A collection of~polymers is called compatible if each distinct pair of polymers within the collection is compatible.
Clearly,  the compatible collections of polymers are in one-to-one correspondence with  the particle configurations.
Finally, let us introduce some notation: We 
write~$\ssP\not\sim\ssP'$ if the polymers~$\ssP$ and~$\ssP'$ 
are not compatible and say that the polymer~$\ssP$ is \emph{in~$\Lambda$} if 
$\ssP\subset \Lambda$.

Let~$\ssP$ be a polymer containing~$N(\ssP)$ sites and occupying both endpoints of~$E(\ssP)$ edges in~$\Z^2$.
We define the Boltzmann weight of~$\ssP$ by the formula
\begin{equation}
%\label{}
\zeta_{\beta,\mu}(\ssP)=e^{\beta E(\ssP)+\mu N(\ssP)}.
\end{equation}
As  is straightforward to verify, the partition function~$\ZG^{\circ,\beta}(\mu,\Lambda)$ can be written as
\begin{equation}
%\label{}
\ZG^{\circ,\beta}(\mu,\Lambda)
=\sum_{\mathscr{P}}\prod_{\ssP\in\mathscr{P}}\zeta_{\beta,\mu}(\ssP),
\end{equation}
where the sum runs over all compatible collections~$\mathscr{P}$ of polymers in~$\Lambda$.

This reformulation of the partition function in the language of compatible polymer configurations allows us to bring 
to bear the machinery of cluster expansion.
Following~\cite{Kotecky-Preiss}, the next key step is a definition of a \emph{cluster}, 
generically denoted by~$\ssC$, by which we will mean a finite non-empty collection of polymers 
that is connected when viewed as a graph with vertices labeled by polymers $\ssP\in\ssC$ and edges connecting pairs of incompatible polymers.  
(Thus, if~$\ssC$ contains but a single polymer it is automatically a cluster. 
If~$\ssC$ contains more than one polymer, 
then any non-trivial division of $\ssC$ into two disjoint subsets has some incompatibility between some pair chosen one from each of the subsets.)
In accord with \cite{Kotecky-Preiss}, a cluster~$\ssC$ is incompatible with a polymer~$\ssP$,
expressed by $\ssC\not\sim\ssP$, if $\ssC\cup\{\ssP\}$ is a cluster.

\smallskip
In order to use this expansion, we need to verify the convergence criterion from~\cite{Kotecky-Preiss}.
In present context this reads as follows: For some $\kappa\ge0$ and any polymer~$\ssP$,
\begin{equation}
\label{KPcond}
\sum_{\ssP'\colon\ssP'\not\sim\ssP}\zeta_{\beta,\mu}(\ssP')e^{(1+\kappa)N(\ssP')}\le N(\ssP).
\end{equation}
Since $\zeta_{\beta,\mu}(\ssP)\le e^{(\mu+2\beta)N(\ssP)}$ is true, this obviously holds if~$\mu$ is sufficiently large 
and negative.
The main result of \cite{Kotecky-Preiss} then says that each cluster~$\ssC$ can be given 
a weight~$\zeta_{\beta,\mu}(\ssC)$ (which is defined less implicitly in \cite{Kotecky-Preiss}), 
such that for all finite volumes~$\Lambda\subset\Z^2$ we have
\begin{equation}
\label{3.27}
\log\ZG^{\circ,\beta}(\mu,\Lambda)=\sum_{\ssC\in\mathscr{C}_\Lambda}\zeta_{\beta,\mu}(\ssC),
\end{equation}
where $\mathscr{C}_\Lambda$ denotes the set of all clusters arising from polymers in~$\Lambda$.
Moreover, this expansion is accompanied by the bound
\begin{equation}
\label{3.28}
\sum_{\ssC\colon\ssC\not\sim\ssP}\bigl|\zeta_{\beta,\mu}(\ssC)\bigr|e^{\kappa N(\ssC)}\le N(\ssP),
\end{equation}
where~$N(\ssC)$ denotes the sum of~$N(\ssP')$ over all~$\ssP'$ constituting~$\ssC$.
With \twoeqref{3.27}{3.28} in hand, we are now ready to prove the first part of the representation 
of~$\ZG^{\circ,\beta}(\mu,\Lambda)$:

\begin{proofsect}{Proof of Lemma~\ref{lemma3.2}}
First, we will introduce a convenient resummation of~\eqref{3.27}.
For each polymer~$\ssP$, let~$\mathscr{N}(\ssP)$ be the set of sites constituting~$\ssP$.
Similarly, for each cluster~$\ssC$, let~$\mathscr{N}(\ssC)$ be the union of~$\mathscr{N}(\ssP)$ over
all~$\ssP$ constituting~$\ssC$.
For each finite~$A\subset\Z^2$, we let 
\begin{equation}
%\label{}
\vartheta_{\beta,\mu}(A)=\sum_{\ssC\colon\mathscr{N}(\ssC)=A}\zeta_{\beta,\mu}(\ssC).
\end{equation}
Clearly,  the weights~$\vartheta_{\beta,\mu}$ are invariant with respect to lattice translations and rotations, having inherited this property from~$\zeta_{\beta,\mu}$.
Moreover, as is easily checked,~$\vartheta_{\beta,\mu}(A)=0$ unless~$A$ is a connected set.
The new weights allow us to rewrite~\eqref{3.27} and~\eqref{3.28} in the following form:
\begin{equation}
\label{3.30}
\log\ZG^{\circ,\beta}(\mu,\Lambda)=\sum_{A\colon A\subset\Lambda}\vartheta_{\beta,\mu}(A),
\end{equation}
with
\begin{equation}
\label{3.31}
\sum_{\begin{subarray}{c}
A\colon 0\in A\\|A|\ge n
\end{subarray}}
\bigl|\vartheta_{\beta,\mu}(A)\bigr|\le e^{-\kappa n}
\end{equation}
for each~$n\ge0$.
Here~$|A|$ denotes the number of sites in~$A$.

Now we are in a position to identify the relevant quantities.
First, the limiting version of the expression~\eqref{3.30} suggests that the pressure should be given by the formula
\begin{equation}
\label{3.32}
\beta \pinfty(\mu)=\sum_{A\colon 0\in A}\frac1{|A|}\vartheta_{\beta,\mu}(A).
\end{equation}
To define the constant~$\tau_1^\circ(\mu)$ representing the wall surface tension, 
let~$\BbbH$ denote the upper half-plane in~$\Z^2$, i.e.,
$\BbbH=\{(x_1,x_2)\in\Z^2\colon x_2>0\}$, and let~$\BbbL$ be the ``line'' in~$\Z^2$ corresponding 
to the boundary of~$\BbbH$, i.e.,~$\BbbL=\{(x_1,x_2)\in\Z^2\colon x_2=0\}$.
Then we define
\begin{equation}
\label{3.33}
\tau_1^\circ(\mu)=-\sum_{\begin{subarray}{c}
A\colon 0\in A\\A\cap\BbbL\ne\emptyset
\end{subarray}}
\frac{|A\cap\BbbH|}{|A|}\,\frac{\vartheta_{\beta,\mu}(A)}{|A\cap\BbbL|}.
\end{equation}
Clearly, in order to contribute to~$\tau_1^\circ(\mu)$, the set~$A$ would have to have 
both~$A\cap\BbbH$ and~$A\setminus\BbbH$ nonempty.
On the basis of \eqref{3.31} it can be shown that the sums in \eqref{3.32} and \eqref{3.33} converge once 
\eqref{KPcond} holds with a $\kappa>0$.

Combining~\eqref{3.30} with~\eqref{3.32}, we can now write that
\begin{equation}
\begin{aligned}
\log\ZG^{\circ,\beta}(\mu,\Lambda)&=
\sum_{x\in\Lambda}\,\,\sum_{\begin{subarray}{c}
A\colon x\in A\\ A\subset\Lambda
\end{subarray}}\,\,
\frac1{|A|}\vartheta_{\beta,\mu}(A)
\\
&=\beta \pinfty(\mu)|\Lambda|-\sum_{x\in\Lambda}
\,\,\sum_{\begin{subarray}{c}
A\colon x\in A\\ A\not\subset\Lambda
\end{subarray}}\,\,
\frac1{|A|}\vartheta_{\beta,\mu}(A).
\end{aligned}
\end{equation}
Using the fact that~$A$ is a connected set and 
thus~$A\cap\Lambda\ne\emptyset$ and~$A\setminus\Lambda\ne\emptyset$ 
imply that~$A\cap\partial\Lambda\ne\emptyset$, the second term on the right-hand side can further be written as
\begin{equation}
\label{3.35}
\begin{aligned}
\quad-\sum_{A\colon A\not\subset\Lambda}&\frac{|A\cap\Lambda|}{|A|}\vartheta_{\beta,\mu}(A)
=-\sum_{x\in\partial\Lambda}\,\sum_{A\colon x\in A}\,
\frac{|A\cap\Lambda|}{|A|}\frac{\vartheta_{\beta,\mu}(A)}{|A\cap\partial\Lambda|}
\\
&\quad=\tau_1^\circ(\mu)|\partial\Lambda|+
\sum_{x\in\partial\Lambda}\,\sum_{A\colon x\in A}\,
\frac1{|A|}\Bigl(\frac{|A\cap\BbbH_x|}{|A\cap\BbbL_x|}-\frac{|A\cap\Lambda|}{|A\cap\partial\Lambda|}\Bigr)
\vartheta_{\beta,\mu}(A).
\quad
\end{aligned}
\end{equation}
Here~$\BbbH_x$ denotes the half-plane in~$\Z^2$ 
that contains~$\Lambda$ and whose boundary $\BbbL_x=\partial \BbbH_x$
includes the portion of the boundary~$\partial\Lambda$ 
that contains~$x$.  (Remember that $\Lambda$ is a rectangular set and thus
its boundary $\partial \Lambda$ splits into four disjoint subsets---the sides of~$\Lambda$.)

Let~$Q_1(\mu)$ denote the (complicated) second term on the right-hand side of~\eqref{3.35}.
Let~$\mathscr{A}$ be the collection of all finite connected sets~$A\subset\Z^2$.
Notice that, whenever a set $A\in \mathscr{A}$ intersects $\partial \Lambda$ in only one of its sides and
$A\cap \partial \Lambda=A\cap\BbbL_x$,  then also $A\cap  \Lambda=A\cap\BbbH_x$, and the corresponding term
in \eqref{3.35} vanishes.
It follows that, in order for the set~$A$ to contribute to the $x$-th term of~$Q_1(\mu)$,
it must contain at least as many sites as is the $\ell^\infty$-distance 
from~$x$ to the sides of~$\partial\Lambda$ not containing~$x$. 
Thus, for a given~$x\in\partial\Lambda$, a set~$A\subset\Z^2$ 
can only contribute to~$Q_1(\mu)$ 
if $A\in\mathscr{A}$ and $|A|\ge \dist(x,\partial\Lambda\setminus\BbbL_x)$.

Since~$|A\cap\Lambda|,|A\cap\BbbH_x|\le|A|$ and~$|A\cap\partial\Lambda|,|A\cap\BbbL_x|\ge1$ 
for any~$A$ contributing to~$Q_1(\mu)$, we can use~\eqref{3.31} to get the bound
\begin{equation}
%\label{}
\bigl|Q_1(\mu)\bigr|\le
\sum_{x\in\partial\Lambda}
\,\,\sum_{\begin{subarray}{c}
A\in\mathscr{A},\,x\in A\\
|A|\ge \dist(x,\partial\Lambda\setminus\BbbL_x)
\end{subarray}}
\bigl|\vartheta_{\beta,\mu}(A)\bigr|\le 
\sum_{x\in\partial\Lambda} 
e^{-\kappa \dist(x,\partial\Lambda\setminus\BbbL_x)}.
\end{equation}
Choosing~$\kappa>0$, letting $G(\kappa)=\sum_{n=1}^\infty e^{-\kappa n}<\infty$, and using $L_1, L_2\in [\theta^{-1} L, \theta L]$ to denote the lengths of the sides of~$\partial \Lambda$, we can bound the right hand side by $8G(\kappa)+2L_1 e^{-\kappa L_2} + 2L_2 e^{-\kappa L_1}$, yielding
$|Q_1(\mu)\bigr|\le8G(\kappa)+4\theta L e^{-\frac{\kappa}{\theta} L}$.
This in turn can be bounded uniformly in~$L$ by a constant that depends only on~$\theta$ 
and we thus get the claim of Lemma~\ref{lemma3.2}.
\end{proofsect}

\subsection{Correlation bounds}
\label{sec3.4}\noindent
This section will be spent on proving Lemma~\ref{lemma3.3}.
We begin by recalling the relevant correlation bounds.
Let us extend our notation~$\langle-\rangle_\Lambda^{\circ,\beta,\mu}$ for the expectation with respect 
to the Gibbs measure in~$\Lambda$ also to the cases when~$\Lambda$ is not necessarily finite.
(It turns out that, by FKG monotonicity, such a state is uniquely defined as a limit of finite-volume Gibbs states 
along any sequence of finite volumes increasing to~$\Lambda$.)
We will use the notation 
\begin{equation}
%\label{}
\langle n_x;n_y\rangle_\Lambda^{\circ,\beta,\mu}=
\langle n_xn_y\rangle_\Lambda^{\circ,\beta,\mu}-
\langle n_x\rangle_\Lambda^{\circ,\beta,\mu}
\langle n_y\rangle_\Lambda^{\circ,\beta,\mu}
\end{equation}
for the truncated correlation function.
This correlation function has the following properties:
\begin{enumerate}
\item[(1)]
For each~$\mu<\mu'\le\mut$ and~$\Lambda\subset\Lambda'$, and all~$x,y\in\Z^2$,
\begin{equation}
%\label{}
\langle n_x;n_y\rangle_\Lambda^{\circ,\beta,\mu}\le
\langle n_x;n_y\rangle_{\Lambda'}^{\circ,\beta,\mu'}.
\end{equation}
\item[(2)]
For each~$\beta>\betac$ there exists a~$\xi=\xi(\beta)<\infty$ such that 
\begin{equation}
%\label{}
0\le\langle n_x;n_y\rangle_\Lambda^{\circ,\beta,\mu}\le e^{-|x-y|/\xi}
\end{equation}
for all  $\mu\le\mut$, all~$\Lambda\subset\Z^2$ and all~$x,y\in\Z^2$.
Here~$|x-y|$ denotes the~$\ell_\infty$ distance between~$x$ and~$y$.
\end{enumerate}
Both~(1) and~(2) are reformulations of well-known properties of the truncated correlation functions for Ising spins.
Namely,~(1) is a simple consequence of the GHS inequality~\cite{GHS}, while~(2) is a consequence of~(1) and the
fact that the infinite-volume truncated correlation function at~$\mu=\mut$ decays exponentially once
$\beta>\betac$. The latter was in turn proved in~\cite{CCS,Bob+Tim}.

\smallskip
A simple consequence of the above observations is the following lemma:

\begin{lemma}
\label{lemma3.4}
Let~$\beta>\betac$. 
Then there exist constants~$\alpha_1=\alpha_1(\beta)\in(0,\infty)$ and~$\alpha_2=\alpha_2(\beta)\in(0,\infty)$ such that 
\begin{equation}
%\label{}
0\le\langle n_x\rangle_{\Lambda'}^{\circ,\beta,\mu}-\langle n_x\rangle_\Lambda^{\circ,\beta,\mu}
\le \alpha_1 e^{-\alpha_2\dist(x,\Lambda'\smallsetminus\Lambda)}
\end{equation}
holds for all~$\mu\le\mut$, all (not necessarily finite) volumes~$\Lambda\subset\Lambda'\subset\Z^2$ and all~$x\in\Lambda$.
\end{lemma}

\begin{proofsect}{Proof}
See, e.g., formula (2.2.6) from \cite{Bob+Tim}; the original derivation goes back to~\cite{Bricmont-Lebowitz-Pfister}.
\end{proofsect}

Now we can start proving Lemma~\ref{lemma3.3}:

\begin{proofsect}{Proof of Lemma~\ref{lemma3.3}}
We begin by a definition of the quantity~$\tau_2^\circ(\mu)$.
Let~$\BbbH$ be the upper half-plane in~$\Z^2$, see Section~\ref{sec3.3}.
Then we define
\begin{equation}
\label{tau2def}
\tau_2^\circ(\mu)=\sum_{\ell\ge1}\bigl(
\langle n_{(0,\ell)}\rangle_{\BbbH}^{\circ,\beta,\mu}-
\langle n_0\rangle_{\Z^2}^{\circ,\beta,\mu}\bigr),
\end{equation}
where~$(x_1,x_2)$ is a notation for a generic point in~$\Z^2$.
By Lemma~\ref{lemma3.4}, the sum converges with a~$\mu$-independent rate (of course, provided~$\mu\le\mut$).

Let~$\Lambda$ be a rectangular volume in~$\Z^2$ with aspect ratio in the interval~$(\theta^{-1},\theta)$.
Let us cyclically label the sides of~$\Lambda$ by numbers~$1,\dots,4$, and define~$\BbbH_1,\dots,\BbbH_4$ 
to be the half-planes in~$\Z^2$ containing~$\Lambda$ and sharing the respective part of the boundary with~$\Lambda$.
Let us partition the sites of~$\Lambda$ into four sets $\Lambda_1,\dots,\Lambda_4$ according to which $\BbbH_j$ 
the site is closest to. We resolve the cases of a tie by choosing the~$\BbbH_j$ with the lowest~$j$.
Now we can write
\begin{multline}
\label{3.41}
\qquad
\langle N_\Lambda\rangle_\Lambda^{\circ,\beta,\mu}-|\Lambda|\langle n_0\rangle_{\Z^2}^{\circ,\beta,\mu}
\\=\sum_{j=1}^4\sum_{x\in\Lambda_j}\bigl(\langle n_x\rangle_\Lambda^{\circ,\beta,\mu}
-\langle n_x\rangle_{\BbbH_j}^{\circ,\beta,\mu}\bigr)
+\sum_{j=1}^4\sum_{x\in\Lambda_j}\bigl(\langle n_x\rangle_{\BbbH_j}^{\circ,\beta,\mu}
-\langle n_0\rangle_{\Z^2}^{\circ,\beta,\mu}\bigr).
\qquad
\end{multline}
If it were not for the restriction~$x\in\Lambda$, the second term on the right-hand side would have the
structure needed to apply~\eqref{tau2def}.
To fix this problem, let~$\BbbS_j$, with~$j=1,\dots,4$, denote the half-infinite slab obtained 
as the intersection~$\BbbH_{j-1}\cap\BbbH_j\cap\BbbH_{j+1}$,
where it is understood that~$\BbbH_0=\BbbH_4$ and~$\BbbH_5=\BbbH_1$.
Clearly,~$\Lambda_j\subset\BbbS_j$ for all~$j=1,\dots,4$.
Then we have
\begin{multline}
\label{3.42}
\qquad\sum_{j=1}^4\sum_{x\in\Lambda_j}\bigl(\langle n_x\rangle_{\BbbH_j}^{\circ,\beta,\mu}
-\langle n_0\rangle_{\Z^2}^{\circ,\beta,\mu}\bigr)
\\
=\tau_2^\circ(\mu)|\partial\Lambda|-
\sum_{j=1}^4\,\sum_{x\in\BbbS_j(\Lambda)\smallsetminus\Lambda_j}\,
\bigl(\langle n_x\rangle_{\BbbH_j}^{\circ,\beta,\mu}
-\langle n_0\rangle_{\Z^2}^{\circ,\beta,\mu}\bigr).
\qquad
\end{multline}
It remains to show that both the first term on the right-hand side of~\eqref{3.41} and the second term
on the right-hand side of~\eqref{3.42} are bounded by a constant independent of~$\mu$ and~$\Lambda$
with the above properties.
As to the first term, we note that, by Lemma~\ref{lemma3.4},
\begin{equation}
%\label{}
\bigl|\langle n_x\rangle_\Lambda^{\circ,\beta,\mu}
-\langle n_x\rangle_{\BbbH_j}^{\circ,\beta,\mu}\bigr|\le 
\alpha_1e^{-\alpha_2\dist(x,\BbbH_j\setminus\Lambda)},
\end{equation}
which after summing over~$x\in\Lambda_j$
gives a plain constant.
Concerning the second contribution to the error, we note that~$\langle n_x\rangle_{\BbbH_j}^{\circ,\beta,\mu}
-\langle n_0\rangle_{\Z^2}^{\circ,\beta,\mu}$ is again exponentially small in~$\dist(x,\Z^2\setminus\BbbH_j)$.
As a simple argument shows, this makes the sum over~$x\in\BbbS_j\setminus\Lambda_j$ finite uniformly 
in~$\Lambda$ with a bounded aspect ratio. 
This concludes the proof.
\end{proofsect}

\section*{Acknowledgments}
\noindent
Part of this paper was written when~M.B. was visiting Center for Theoretical Study in Prague.
The research of R.K. was partly supported by
the grants GA\v{C}R~201/03/0478 and MSM~110000001.
The research of L.C.~was supported by the NSF under the grant DMS-9971016 and
by the NSA under the grant NSA-MDA~904-00-1-0050.~R.K. would also like to thank 
the Max-Planck Institute for Mathematics in Leipzig for their hospitality
as well as the A.~von Humboldt Foundation whose Award made the stay in Leipzig possible.

\end{document}